Force-free state in a superconducting single crystal and angle-dependent vortex helical instability.

J. del Valle[1,(a)], A. Gomez[1, (b)], E. M. Gonzalez[1,2], S. Manas-Valero[3], E. Coronado[3], J. L. Vicent[1,2]

[1] Departamento Física Materiales, Facultad CC. Fisicas, Universidad Complutense, 28040 Madrid (Spain)

[2] IMDEA-Nanociencia, Faraday 8, Cantoblanco, 28049 Madrid (Spain)

[3] Instituto de Ciencia Molecular (ICMol), Universitat de Valencia, Catedratico Jose Beltran 2, 46980 Paterna (Spain)

*Abstract:*

Superconducting 2H-NbSe$_2$ single crystals show intrinsic low pinning values. Therefore, they are ideal materials with which to explore fundamental properties of vortices. (V, I) characteristics are the experimental data we have used to investigate the dissipation mechanisms in a rectangular shape 2H-NbSe$_2$ single crystal. Particularly, we have studied dissipation behavior with magnetic fields applied in the plane of the crystal and parallel to the injected currents, i.e. in the force-free state where the vortex helical instability governs the vortex dynamics. In this regime, the data follow the elliptic critical state model and the voltage dissipation shows an exponential dependence, $V \propto e^{\alpha(I-I_{c||})}$, $I_{c||}$ being the critical current in the force-free configuration and $\alpha$ a linear temperature dependent parameter. Moreover, this exponential dependence can be observed for in-plane applied magnetic fields up to 40º off the current direction, which implies that the vortex helical instability plays a role in dissipation even out of the force-free configuration.

(a) Present address: Department of Physics, University California-San Diego, CA 92093 USA.
(b) Present address: Centro de Astrobiologia, INTA-CSIC, 28850 Torrejon de Ardoz, Spain

Vortex physics is a long-standing topic which remains open in many scenarios. Vortices are objects which can be entangled, crossed, cut and reconnected. The study of vortices as topological defects encompasses many fields, such as plasmas [1], fluids [2-5], superfluids [6-9], and superconductors [10-15]. Sometimes fundamental properties of superconducting vortices are blurred by the crucial role played by vortex pinning. To avoid this problem and capture basic vortex dynamics in superconductors, materials with very low pinning are required; for example, superconducting dichalcogenide single crystals [16]. These layered platelets present large crystal sizes, they are very stable and very easy to handle too. They show very low Ginzburg number ($Gi$) ~ $10^{-6}$; that is very low thermal fluctuations and very low critical current density ($J_C$) in comparison with the depairing current density ($J_0$), for instance $J_C / J_0$ ~ $10^{-5}$, thus showing very low pinning. These crystals are an ideal tool to explore fundamental superconducting properties. Some topics which have been studied: 2D superconductivity; i. e. superconducting samples whose thickness is much lower than Ginzburg-Landau coherence length [17]; perfect Abrikosov vortex lattice [18]; enhancement of the critical current slightly below the upper critical field $H_{c2}$ (peak effect) [19]; static and dynamic vortex lattices [20]; direct imaging of vortex lattice structures during creep [21]; surface barriers and edge current enhancement [22]; Fermi surface sheet-dependent superconductivity (multiband superconductivity) [23]; geometrical phase transitions in mesoscopic superconductors [24]; coexistence of stable vortex phases [25]; interplay between charge density waves and superconductivity [26]; superconductivity with atomic-scale thickness [27]; vortex dynamics in superconducting nanowires [28]; Ising-superconductivity in a monolayer [29]. In summary, single crystal dichalcogenides have been ideal materials to study many superconducting topics.

In this paper, we report on the vortex lattice dynamics in the force-free state in single crystal dichalcogenide 2H-NbSe$_2$. In this force-free state the magnetic field is applied in-plane and parallel to the electrical current direction. Consequently, in theory, the Lorentz force on the vortex lattice is null. Therefore, from the theoretical point of view and in pin-free situation the supercurrent can flow infinitely. In 1966, Josephson argued that the superconducting mixed state allows unstable states to be realized. He pointed out that, if the sample is carrying a current in a longitudinal magnetic field (force-free configuration), the vortices can develop a helical geometry and cutting of the flux lines can occur [30]. We have to note that recently, in layered superconductors, helical instability has been proposed [31], but the origin and the magnetic field and electrical current configurations of this helical instability are fully different from force-free configuration.

The main source of dissipation of a superconductor in the force-free configuration is the helical instability studied by Clem for a single vortex [32] and for a vortex lattice by Brandt [33]. The origin of this instability is in the own transport current generating self-magnetic fields. The vortex is unstable to spiral perturbations, thus, increasing the transport current, clockwise spirals enhance while counter-clockwise spirals weaken, and flux-line cutting processes can occur.

Since these pioneering works, force-free state has gained the attention of many researchers and currently the topic is still open; see for example, the review by Campbell, and references therein [34].

In the present work, dissipation and vortex dynamics are studied in the force-free state, by means of (V, I) characteristic curves measured in 2H-NbSe$_2$ single crystal. We have found that the (V, I) curves follow an exponential law in the force-free state and this trend survives for applied magnetic field directions far from the force-free configuration.

2H-NbSe$_2$ single crystals have been grown by the usual method of iodine-vapor transport from stoichiometric pre-reacted powders sintered at 900° C [35]. The single-crystal growth details can be found in [36]. A rectangular shape sample was obtained following exactly the same method used by Xiao et al [37]. These authors studied (I, V) curves in 2H-NbSe$_2$ large single crystal; which was cut by hand using a razor blade. In our case, the final rectangular sample is 8 (length) x 2 (wide) x 0.15 (thickness) mm$^3$. Low resistance silver paint was used for the four contacts. The separation between the voltage contacts was 1 mm. The sample was mounted in a rotatable sample holder in a commercial cryostat with superconducting solenoid. (V, I) characteristics were measured with standard four terminal method. The electrical field criterion ($E_{crit}$) for critical current was: $E_{crit}$ = 0.05 µV mm$^{-1}$. The zero magnetic field critical temperature is 7.2 K and $\Delta T_c$ = 75 mK (see Fig. 1). Inset in Fig. 1 shows M(H) measurements at 4.2 K. From these data a critical current density $J_c(0)$ = 1.6 x 10$^4$ A cm$^{-2}$ is obtained using the width of the magnetization curve [38]. This value is similar to reported values, see for instance [39]. The magnetic field was applied always in plane, as can be seen in Fig. 2, which shows a sketch of the experimental setup configuration. In the measurements we have followed the usual protocol, after cool down the sample to the selected temperature, the magnetic field is applied and finally the direct electrical current is injected.

First of all, we have compared our experimental data with the elliptic critical state model (ECSM) [40]. The main feature of this model is that the critical current ($I_C$) is dependent on both perpendicular and parallel components of the current ($I_\perp$, $I_{||}$). According to the formula:

$$I_C(\varphi) = I_{C\perp} \frac{1}{\sqrt{sin^2(\varphi) + \left(\frac{I_{C\perp}}{I_{C||}} cos(\varphi)\right)^2}} \quad (1)$$

This behavior promotes flux cutting processes. Critical currents have been measured as a function of φ and then, they have been decomposed into the perpendicular [$I_{C\perp}(\varphi) = I_C(\varphi) sin(\varphi)$] and parallel [$I_{C||}(\varphi) = I_C(\varphi) cos(\varphi)$] components to the applied magnetic field. Fig. 3 shows the normalized critical currents in terms of their perpendicular and parallel components. The line represents a fit to the ECSM model, being $I_{C\perp} = I_C(90°)$ and $I_{C||} = I_C(0°)$. The experimental data follow the ECSM for different applied magnetic fields, checking the validity of the model.

Next we study the temperature dependence of the force-free state. Fig. 4(a) shows (V, I) characteristic curves with an applied magnetic field of 1 T. We observe that the helical instability and flux cutting dynamics lead to an exponential regime above $I_{C||}(T)$, before reaching the flux flow situation.

$$V \propto e^{\alpha(I - I_{C||})} \quad (2)$$

The same behavior has been obtained in applied magnetic fields from 0.1 T to 6T. Using a linear fit, the slope of the log (V) *vs*. I curves has been obtained, see Fig. 4(b). This slope is equal to the exponent $\alpha$, which controls how fast dissipation increases with the current after $I_{d|}$ is surpassed. Fig. 4(b) shows the temperature dependence of $\alpha$ :

$$\alpha \propto \left(1 - \frac{T}{T_C}\right) \tag{3}$$

as can be seen, $\alpha$ grows as the temperature is decreased.

Finally, we measure and analyze the angular dependence of the vortex helical instability. Fig. 5 shows the (V, I) curves for H=1 T, varying the in-plane angle $\varphi$ between H and I at constant temperature. Note that the voltage axis is plotted in logarithmic scale. The linear voltage response (V~I) characteristic of the flux flow regime for the usual Lorentz force configuration ($\varphi=90°$) is seen as logarithmic shape in this representation. Another experimental fact that has to be underlined is that $\varphi$ = 0º (force–free state) and $\varphi$ = 10º are very similar. Once the usual Lorentz forces are acting, i. e. $\varphi$ > 0º, we can analyze the competition between the two dissipation contributions, Lorentz force and helical instability, if any. The analysis of these data is done plotting the derivative of the log (V)/I *vs.* I (the injected electrical currents). If the (V, I) curve follows an exponential law, the derivative will be constant. Fig. 6 shows these plots for different values of $\varphi$. The data are extracted from the raw (V, I) characteristic curves of Fig. 4. We observe that up to $\varphi$ = 40º and, for a particular interval of input currents, a constant derivative value is obtained. Therefore, in this regime, the output voltages show an exponential law, so the helical instability dominates the dissipation processes. Increasing further the driving currents the usual flux flow develops.

In summary, we have found that measuring (V,I) characteristic curves in low pinning 2H-NbSe$_2$ single crystals we were able to obtain the fundamental physics of vortex matter, as are dissipation mechanisms in the force-free state. The main findings of this work are: i) In the force-free state configuration the voltage dissipation follows an exponential law, induced by the vortex helical instability. Increasing the magnitude of the injected current an exponential response is found in the output voltage. If the driving current is increased further the usual linear flux flow appears; ii) For magnetic fields applied in-plane, but out of the injected current direction, the helical instability state remains active and it can be detected up to 40º off the current direction; iii) The experimental data accomplish the elliptic critical state model.

We are grateful for support from Spanish Ministerio de Economia, Industria y Competitividad Grants No. FIS2013-45469-C4-1-R and FIS2016-76058-C4-1-R (AEI/FEDER,EU) and Comunidad de Madrid Grant No. P2013/MIT-2850. E.C. and S.M.-V. acknowledge the financial support from the Spanish Ministerio de Economia, Industria y Competitividad (MAT2014-56143-R co-financed by FEDER and Excellence Unit "María de Maeztu" MDM-2015-0538) and the Generalitat Valenciana (Program Prometeo). S.M.-V. thanks the Spanish Ministerio de Economia, Industria y Competitividad for a F.P.U. fellowship (FPU14/04407).

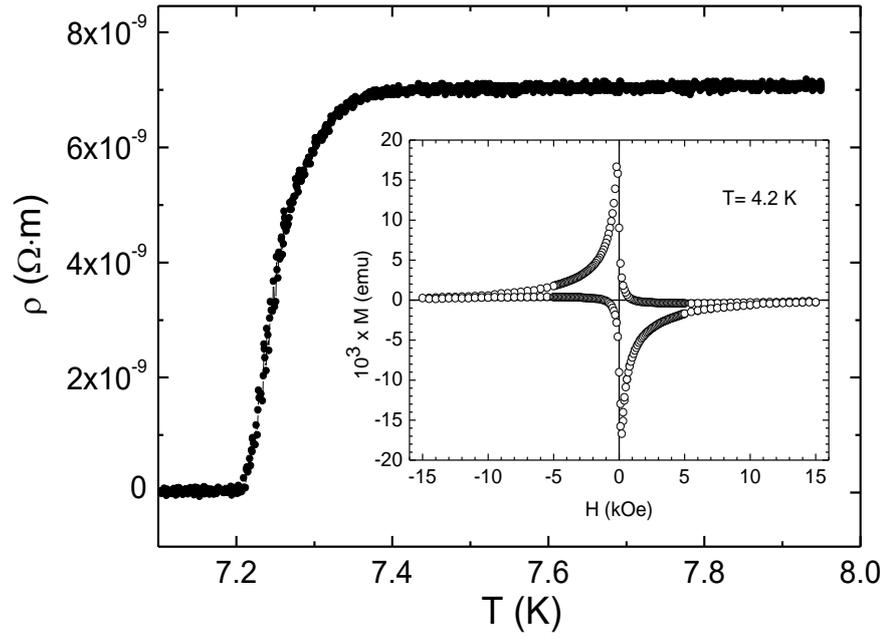

FIG. 1. 2H-NbSe$_2$ single crystal superconducting resistivity transition (critical temperature $T_c$ = 7.2 K). Inset shows M (H) hysteresis loop at 4.2 K (critical current density $J_c$ (H = 0) = 1.6 x 10$^4$ A cm$^{-2}$).

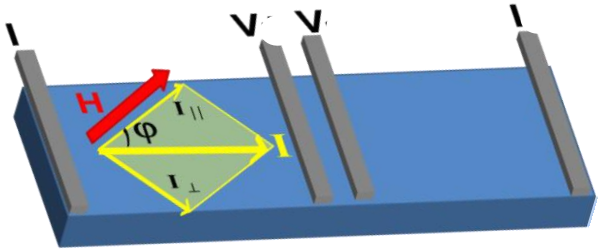

FIG. 2. Sketch of the measurement configuration. Both current (yellow) and magnetic field (red) are applied parallel to the NbSe$_2$ planes (blue). The angle φ between them is varied from 90° (maximum Lorentz force) to 0° (force-free configuration), using a rotatable sample holder.

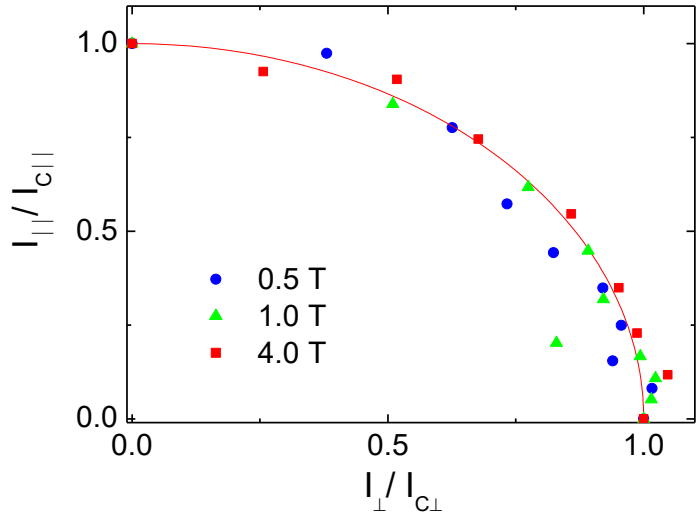

FIG. 3. Parallel (Y-axis) and perpendicular (X-axis) components of $I_c(\varphi)$ normalized to $I_c(0º)$ and $I_c(90º)$, respectively, for three different magnetic fields: 0.5 T (dots), 1 T (triangles) and 4 T (squares). Line is a fit to the ECSM model (see text).

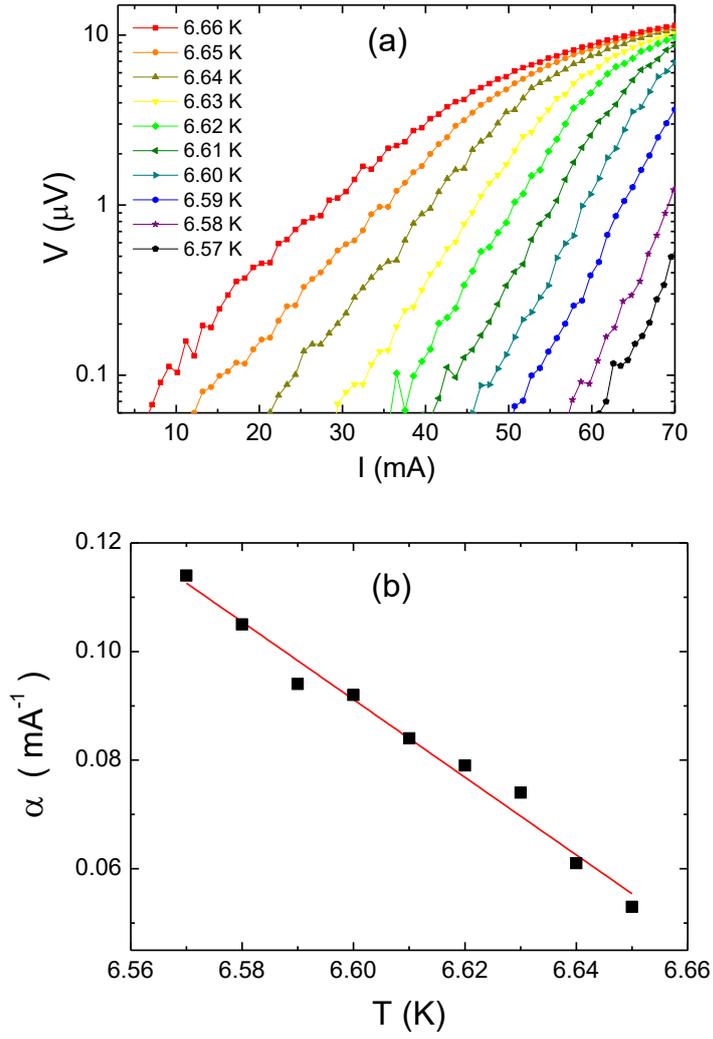

FIG. 4. (a) V-I characteristics for $\mu_0 H=1T$ and $\varphi=0°$ (force-free state) for different temperatures. V is in logarithmic scale. (b) Slope of the logV - I curves as a function of the temperature. The linear dependence is shown by a linear fitting (red line).

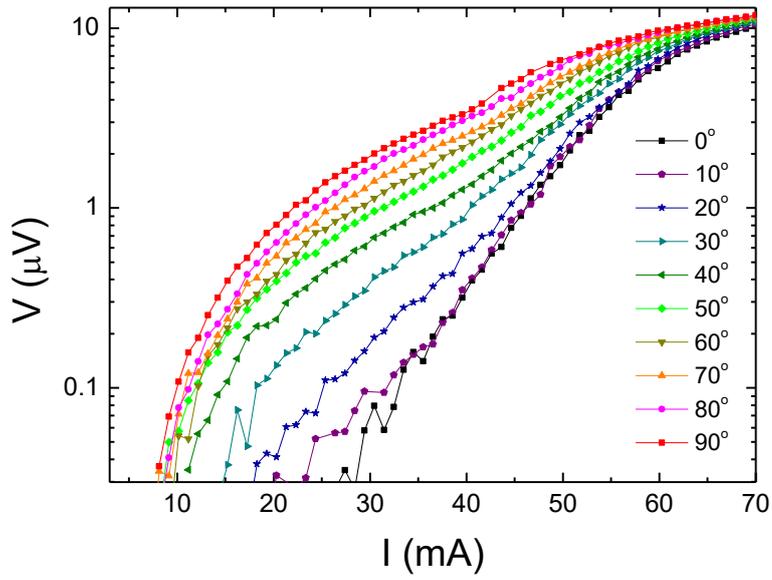

FIG. 5. V-I characteristics for $\mu_0 H=1T$ and T = 6.63 K. Upper curve is for $\varphi$ = 90º (usual Lorentz force configuration). The lower curve is for $\varphi$ = 0º (force-free state configuration). The separation between two sequential curves is 10º.

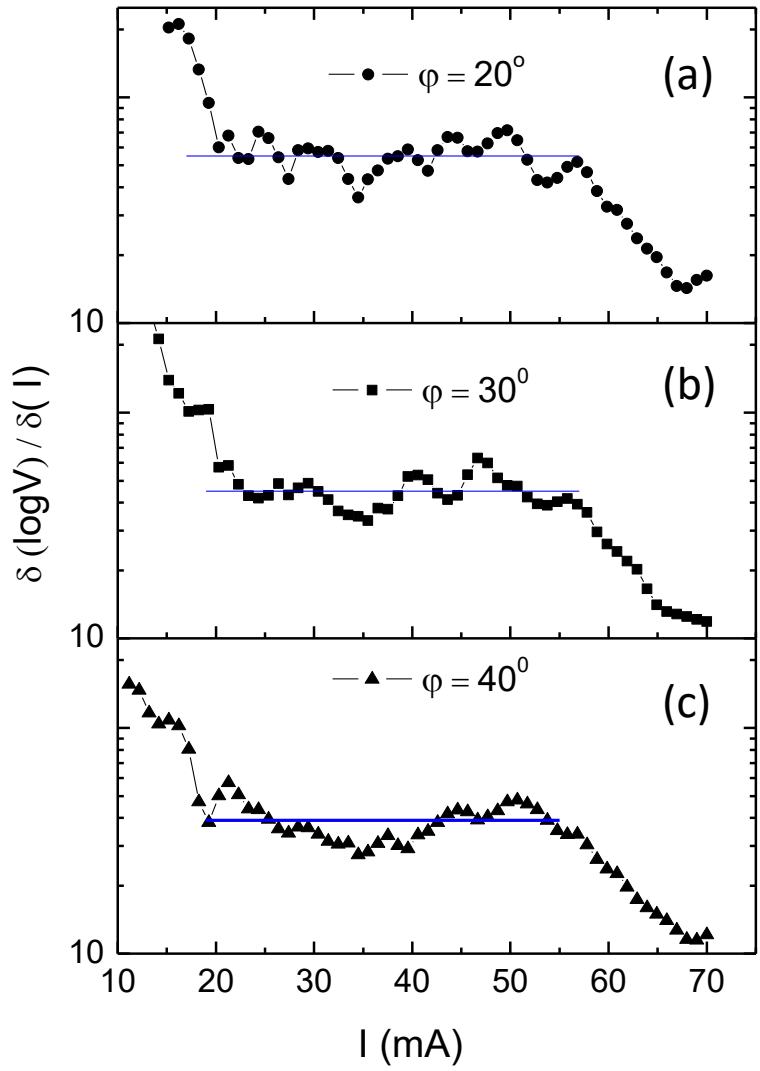

FIG. 6. Derivative of the log V-I curves vs. I for different angles between the applied magnetic field and the applied current direction: (a) φ=20°, (b) φ=30°, (c) φ=40°. The horizontal lines are guides to the eye.